\documentclass[runningheads]{llncs}

\usepackage{graphicx,booktabs,tabularx,xcolor,hyperref}
\usepackage{amsmath,subcaption,float,amsfonts,amssymb,bm,mathtools,booktabs,threeparttable,nccmath}
\usepackage[title]{appendix}

\hypersetup{colorlinks=true,linkcolor=[rgb]{0.1,0.3,0.9},urlcolor=[rgb]{0.2,0.2,0.2},citecolor=[rgb]{0.1,0.3,0.9}}

\makeatletter\newcommand*\mysizea{\@setfontsize\mysizea{9.5}{11}}\makeatother
\makeatletter\newcommand*\mysizeb{\@setfontsize\mysizeb{8.5}{11}}\makeatother
\makeatletter\newcommand*\mysizec{\@setfontsize\mysizec{7}{11}}\makeatother
\makeatletter\newcommand*\mysized{\@setfontsize\mysized{6.5}{11}}\makeatother

\begin{document}

\title{Non-Reference Quality Assessment for Medical Imaging: Application to Synthetic Brain MRIs}

\titlerunning{Non-Reference Medical Image Quality Assessment}

% \author{Anonymous authors}
% \institute{Anonymous organisation \\ \email{anonymous@anonymous}}
% \authorrunning{Anonymous author et al.}

\author{Karl \small{Van} Eeden Risager \and Torkan Gholamalizadeh \and Mostafa Mehdipour Ghazi}
\institute{Pioneer Centre for AI, Department of Computer Science, University of Copenhagen \\
Research and Development, 3Shape A/S, Copenhagen, Denmark \\ \email{ghazi@di.ku.dk}}
\authorrunning{Risager et al.}

\maketitle

\begin{abstract}

Generating high-quality synthetic data is crucial for addressing challenges in medical imaging, such as domain adaptation, data scarcity, and privacy concerns. Existing image quality metrics often rely on reference images, are tailored for group comparisons, or are intended for 2D natural images, limiting their efficacy in complex domains like medical imaging. This study introduces a novel deep learning-based non-reference approach to assess brain MRI quality by training a 3D ResNet. The network is designed to estimate quality across six distinct artifacts commonly encountered in MRI scans. Additionally, a diffusion model is trained on diverse datasets to generate synthetic 3D images of high fidelity. The approach leverages several datasets for training and comprehensive quality assessment, benchmarking against state-of-the-art metrics for real and synthetic images. Results demonstrate superior performance in accurately estimating distortions and reflecting image quality from multiple perspectives. Notably, the method operates without reference images, indicating its applicability for evaluating deep generative models. Besides, the quality scores in the [0, 1] range provide an intuitive assessment of image quality across heterogeneous datasets. Evaluation of generated images offers detailed insights into specific artifacts, guiding strategies for improving generative models to produce high-quality synthetic images. This study presents the first comprehensive method for assessing the quality of real and synthetic 3D medical images in MRI contexts without reliance on reference images.

\end{abstract}


\section{Introduction}

Generating synthetic brain MRI images using deep learning addresses key challenges in medical imaging \cite{frid2018synthetic}, such as domain adaptation and the scarcity of labeled data. Synthetic images enhance model robustness and generalizability across clinical settings with variations in imaging protocols, scanner types, and patient populations \cite{kamnitsas2017unsupervised}, and mitigate privacy concerns by providing anonymized data compliant with privacy data sharing regulations in medical research \cite{yi2019generative}.

High-quality synthetic images are essential for the reliability of deep learning tools in clinical diagnosis and treatment planning \cite{gao2021generating}. Recent advancements in deep generative models, such as denoising diffusion probabilistic models (DDPM) \cite{ho2020denoising}, have shown potential in creating realistic samples in comparison with traditional generative adversarial networks (GANs) \cite{goodfellow2014generative} and variational autoencoders (VAEs) \cite{kingma2013auto}. However, these methods often have high computational costs when applied to high-resolution 3D brain MRIs. The 3D wavelet diffusion model (WDM) \cite{friedrich2024wdm} has recently addressed high-dimensional data challenges, showing promising results that outperform state-of-the-art techniques. Despite these advancements, comprehensive quality assessment of images generated by these models remains limited, which is crucial for meeting clinical standards and the trustworthiness of synthetic data.

Existing image quality metrics like the structural similarity index measure (SSIM) \cite{wang2004image} and peak signal-to-noise ratio (PSNR) \cite{hore2010image} require reference images, posing challenges for generative models without direct one-to-one correspondence. Metrics like Fréchet inception distance (FID) \cite{obukhov2020quality} evaluate similarity based on groups of images, not individual quality. Traditional non-reference metrics such as the blind image spatial quality evaluator (BRISQUE) \cite{mittal2012no} and perception-based image quality evaluator (PIQE) \cite{venkatanath2015blind} or learning-based blind metrics using neural image assessment (NIMA) \cite{talebi2018nima} and generated image quality assessment (GIQA) \cite{gu2020giqa} are designed for 2D natural images and are less effective for the complexities of medical imaging.

These metrics fail to capture specific artifacts inherent in MRI modalities \cite{erasmus2004short}, such as motion artifacts, bias field, and complex noise \cite{gudbjartsson1995rician}. There is a need for quality assessment methods tailored to medical imaging that account for domain-specific variations and ensure high-quality synthetic images. To this end, we propose six distinct metrics to evaluate MRI quality by addressing common artifacts. We train a 3D ResNet model based on a pre-trained 2D ResNet-50 \cite{he2016deep} to estimate image quality concerning these artifacts. This comprehensive, non-reference image quality assessment is validated using several public datasets and compared against state-of-the-art metrics.

Moreover, we train a diffusion model \cite{friedrich2024wdm} on various brain MRI datasets to generate high-quality synthetic 3D images and assess their quality. Our evaluation results demonstrate precise quality estimation for different aspects of both real and synthetic images, and the proposed quality assessor provides a more accurate reflection of image quality in synthetic samples generated by the generative model, which can help enhancements in the quality of generated images, such as by adjusting loss functions to penalize specific artifacts. To the best of our knowledge, this is the first time a comprehensive method has been proposed for 3D medical imaging quality assessment in the context of MRIs.

\section{Methods}

We train a 3D deep regression network designed to assess specific quality aspects of MRI scans through multiple output responses. Using high-quality reference images as input, we employ dynamic data augmentation techniques to simulate various distortions \cite{ghazi2022fast}. These distortions, representing common MRI artifacts, are evaluated using metrics scaled from 0 to 1, where 0 and 1 denote the lowest and highest image quality, respectively.

During training, distortions are randomly applied to the reference image in each iteration, varying in strength. This method exposes the network to a wide range of distortion levels, thereby enhancing its robustness and accuracy in quality estimation under diverse testing conditions. By incorporating diverse artifacts during training, the network effectively learns to generalize across different distortions, ensuring reliable performance in real-world scenarios.

\subsection{Generative Network}

To generate synthetic 3D brain MRIs of high quality, we employ the Wavelet diffusion model (WDM) \cite{friedrich2024wdm}. This approach utilizes a discrete wavelet transform (DWT) to decompose input images into eight wavelet coefficients at half the original spatial resolution. These coefficients are then used as inputs to a denoising diffusion probabilistic model \cite{ho2020denoising}, which iteratively diffuses noise into the transforms and denoises them using a Markov chain process. The model is optimized on the denoised coefficients, and finally, an inverse DWT is applied to the denoised coefficients to generate synthetic images at full resolution.

\subsection{Quality Metrics}

\subsubsection{Contrast Change}

To measure the contrast deviation in the modified image $J$ from the high-contrast reference image $I$, we use the standard deviation ratio (SDR) or root mean square contrast, defined as $\sigma_J / \sigma_I$, where $\sigma$ represents the intensity standard deviation \cite{peli1990contrast}. We simulate various contrast levels by applying a gamma transform $J = I^\gamma$ \cite{chen2019multi} with random $\gamma$ values in the range [0.5, 2]. For $\gamma > 1$, we ensure SDR remains within [0, 1] by using $\sigma_I / \sigma_J$.

\subsubsection{Bias Field}

We quantify the bias in the nonuniform image $J$ relative to the homogeneous reference image $I$ using the coefficient of variation ratio (CVR), defined as $\sigma_I \mu_J / \sigma_J \mu_I$, where $\mu$ denotes the average intensity. To simulate intensity variation across the image, we multiply an elliptic gradient field with the brain volumes \cite{hui2010fast}. This gradient field is computed using the equation of an ellipse in standard form, with points from a structured rectangular grid ranging from 1 to 224, random centers within [1, 224], and radii set to 224.

\subsubsection{Gibbs Ringing}

The ringing effect in the distorted image $J$ is measured using the truncation ratio of high-frequency components, defined as $f_c / 224$, where $f_c$ is the cutoff frequency of the fast Fourier transform (FFT) of the ringing image, with the same cubic dimension of 224. To simulate this artifact, we apply the centralized FFT (k-space) to the brain volumes in three orthogonal directions and truncate the edges of the k-space \cite{moratal2008k} at a random cutoff frequency $f_c$ within the range [32, 224] along each of the three axes.

\subsubsection{Motion Ghosting}

We measure the motion effect in the ghosting image $J$ compared to the clean reference image $I$ using the modulation factor of the k-space lines, defined as $\alpha = \min(F_I / F_J)$, where $F$ represents the image FFT. To simulate this artifact, we randomly weigh k-space lines with a random factor $\alpha$ within the range [0.35, 1] along the three axes \cite{moratal2008k}.

\subsubsection{Rician Noise}

To quantify noise strength in the noisy image $J$ relative to the noise-free reference image $I$, we employ the PSNR, defined as $\mathrm{PSNR} = 10 \log_{10} (1 / \mathrm{MSE})$, where $\mathrm{MSE} = \sum_{i=1}^{N} (I_i - J_i)^2 / N$ represents the mean squared error between $I$ and $J$ with $N$ pixels. Rician noise is introduced by adding zero-mean Gaussian noise to the real and imaginary components of the image, and the noise level is randomly selected from [10$^{-6}$, 10$^{-2}$]. To maintain the PSNR within [0, 1], the score is downscaled by 100 and clipped if it exceeds 1, corresponding to minimal noise levels.

\subsubsection{Blur Effect}

The blurriness of the image $J$ relative to the sharp reference image $I$ is quantified using the ratio of high-frequency components, defined as $|F_J > T_J| / |F_I > T_I|$, where $|F > T|$ denotes the count of high-frequency components exceeding threshold $T$ in the k-space. Thresholds are set to one-thousandth of the maximum frequency in the k-spaces \cite{de2013image}. To simulate blur effects, we employ two random approaches: resampling and Gaussian smoothing. For resampling, images are downsampled/upsampled to their original dimensions using linear interpolation with a random scaling factor in [0.2, 2]. For smoothing, a Gaussian filter with a random kernel size in [3, 11] and sigma in [0.25, 5] is applied.

\subsection{Quality Network}

Based on the distortions applied and evaluated on the fly, we train a 3D ResNet-50 architecture with six regression fully connected outputs, each estimating the image quality from a specific artifact viewpoint. The network is trained using full-size brain MRIs of isotropic dimension 224 $\times$ 224 $\times$ 224. Since the responses are in the range [0, 1], we apply a Sigmoid activation layer in the output. Initially, we used MSE loss with a large mini-batch size for training the network, but we observed a tendency to neglect responses with lower-quality scores. Therefore, we decreased the mini-batch size and employed a focal MSE loss. The proposed focal MSE loss is defined as
\begin{equation*}
\mathcal{L} = \frac{1}{M} \sum_{m=1}^{M} \big(1 + \alpha |Y_m - T_m| ^ \gamma \big) \big(Y_m - T_m \big) ^ 2 \,,
\end{equation*}
where $M = 6$ is the number of output responses in the network, $Y$ and $T$ represent the predicted and ground-truth target values, and $\alpha$ and $\gamma$ are the focal loss parameters, experimentally set to 2 and 1, respectively.

\subsection{Augmentation and Inference}

To make the predictions robust to geometrical and shape variations, we randomly apply various augmentations on the fly, including translations in the range of [-10, 10] pixels along the three axes, rotations in the range of [-10$^\circ$, 10$^\circ$] about the three axes, plane flipping, elastic deformations \cite{simard2003best}, and skull stripping. Skull stripping is performed by randomly cropping around the brain using morphological operations such as dilation and erosion, and elastic deformations are obtained by interpolating a random 3D uniform displacement field, which is smoothed using a Gaussian filter with $\sigma$ in the range [20, 30], scaled by $\alpha$ in the range [200, 500], and added to a structured grid with values from 1 to 224. These augmentations enable the network to learn representations from brain MRIs for quality assessment, regardless of the presence or shape of objects and their geometrical variations. Finally, we apply flipping and average the prediction scores from the original and flipped images to obtain more robust quality scores during inference.

\section{Experiments and Results}

\subsection{Data}

The datasets utilized in this study comprise a variety of T1-weighted brain MRIs acquired from different scanners (Siemens, GE, and Philips) operating at varying field strengths (1.5T and 3T). Specifically, the datasets include 503 high-quality (distortion-free) reference images from ADNI \cite{wyman2013standardization} preprocessed using FreeSurfer \cite{fischl2002whole}; 40 MRI scans from OASIS \cite{marcus2007open}; 30 MRIs from Hammers \cite{faillenot2017macroanatomy}; 18 images from IBSR \cite{rohlfing2011image}; 581 scans from IXI \cite{rossi2019analysis}; 180 MRIs from SynthRAD \cite{thummerer2023synthrad2023}; and 1,252 brain MRIs from BraTS \cite{baid2021rsna}. All the image intensities are normalized to the range [0, 1] based on the data type stored in each NIFTI file, resampled to an isotropic resolution of 1 mm, and padded and center-cropped to 224 $\times$ 224 $\times$ 224, to standardize the training process. % Additional details regarding the datasets and preprocessing methods are available in the supplementary document.
Additional details regarding the datasets and preprocessing methods are available in Appendix \ref{appendix-data}.

\subsection{Setup}

We utilized ADNI, BraTS, and SynthRAD datasets to train, validate, and test the WDM generative network. Training followed hyperparameters recommended in \cite{friedrich2024wdm}, running for up to $6 \times 10^{5}$ iterations, with validation based on FID using VGG16 \cite{simonyan2014very} as a feature extractor. Additionally, the ADNI dataset was utilized exclusively for training, validating, and testing the 3D ResNet-50 quality network, while the remaining datasets were reserved solely for evaluation purposes. For the quality network, we set the base learning rate and weight decay to $10^{-4}$ while optimizing the network using the Adam algorithm across 50 epochs, implementing a piecewise learning rate schedule with a drop factor of 0.9, a gradient decay factor of 0.9, and a squared gradient decay factor of 0.99.

\subsection{Results}

Initially, we assessed the generalization capability of the ADNI-trained quality network on the real ADNI test set by introducing various random distortions to the reference images, including individual and mixed artifacts. Table \ref{table-adni-test-real} summarizes the results of this experiment. The predicted quality values closely match the ground-truth values, indicating minimal estimation error. However, notable discrepancies are observed especially in cases involving blur effects, where interpolation and smoothing artifacts may lead to confusion with other distortions.

Next, we evaluated the generalization performance of the quality network on external test sets, with and without random distortions including mixed artifacts, and compared its performance against state-of-the-art image quality assessors. Table \ref{table-test-real} presents the results of this experiment. The proposed network demonstrates superior accuracy in estimating image quality across diverse datasets compared to other metrics like PIQE, especially given the high quality of the original preprocessed images. Notably, the absence of results in certain cells underscores the limitations of SSIM and PSNR, which necessitate reference image pairs for assessment. Additionally, the interpretability of BRISQUE scores is challenging due to its undefined score range. These findings highlight the network's ability to effectively assess MRI scan quality for preprocessed and distorted images without reliance on reference images.

\begin{table*}[!b]
\centering
\mysized
\caption{Comparison of the quality scores and errors (mean$\pm$SD) estimated using the proposed method for different distortions applied to real ADNI test samples.}
\label{table-adni-test-real}
\renewcommand{\arraystretch}{1.2}
\centering
\begin{tabular}{lccccccc}
\toprule
 & Contrast & Bias & Ring & Ghost & Noise & Blur & Mixed \\
\bottomrule
Ground-truth & 0.771$\pm$0.114 & 0.915$\pm$0.078 & 0.553$\pm$0.242 & 0.625$\pm$0.188 & 0.466$\pm$0.074 & 0.605$\pm$0.216 & 0.699$\pm$0.219 \\
Predicted & 0.796$\pm$0.145 & 0.960$\pm$0.007 & 0.570$\pm$0.247 & 0.646$\pm$0.185 & 0.513$\pm$0.136 & 0.691$\pm$0.242 & 0.732$\pm$0.214 \\
MSE & 0.009$\pm$0.016 & 0.007$\pm$0.019 & 0.010$\pm$0.022 & 0.003$\pm$0.004 & 0.009$\pm$0.024 & 0.010$\pm$0.009 & 0.011$\pm$0.023 \\
\toprule
\end{tabular}
\end{table*}

\begin{table*}[!t]
\centering
\mysizeb
\caption{Comparison of quality assessment scores (mean$\pm$SD) using different methods across various real datasets.}
\label{table-test-real}
\renewcommand{\arraystretch}{1.2}
\centering
\begin{tabular}{lccccc}
\toprule
 & Proposed & 1 - PIQE & SSIM & BRISQUE & PSNR \\
\bottomrule
ADNI & 0.955$\pm$0.017 & 0.580$\pm$0.065 & &
38.285$\pm$5.212 & \\
Distorted & 0.732$\pm$0.070 & 0.267$\pm$0.191 & 0.545$\pm$0.105 & 43.846$\pm$7.175 & 17.821$\pm$2.242 \\
\bottomrule
IXI & 0.955$\pm$0.025 & 0.663$\pm$0.051 & & 42.902$\pm$2.290 & \\
Distorted & 0.699$\pm$0.064 & 0.220$\pm$0.217 & 0.567$\pm$0.118 & 44.780$\pm$5.405 & 22.577$\pm$3.437 \\
\bottomrule
OASIS & 0.961$\pm$0.006 & 0.568$\pm$0.058 & & 42.931$\pm$0.790 & \\
Distorted & 0.719$\pm$0.066 & 0.211$\pm$0.187 & 0.588$\pm$0.121 & 44.899$\pm$4.670 & 21.823$\pm$2.857 \\
\bottomrule
Hammers & 0.945$\pm$0.020 & 0.311$\pm$0.057 & & 45.373$\pm$0.736 & \\
Distorted & 0.699$\pm$0.072 & 0.154$\pm$0.183 & 0.654$\pm$0.147 & 45.515$\pm$3.364 & 27.209$\pm$3.234 \\
\bottomrule
IBSR & 0.935$\pm$0.013 & 0.226$\pm$0.041 & & 44.030$\pm$1.651 & \\
Distorted & 0.701$\pm$0.081 & 0.131$\pm$0.190 & 0.670$\pm$0.158 & 46.380$\pm$3.368 & 25.633$\pm$3.063 \\
\toprule
\end{tabular}
\end{table*}

Furthermore, to assess the quality of images generated by deep generative models, we compared the quality of different generated datasets using our proposed method with state-of-the-art metrics. Table \ref{table-test-fake1} displays the results of this comparative study. The scores obtained through our method align with SSIM, BRISQUE, PSNR, and FID scores. Notably, our proposed scores provide a more interpretable and comprehensive understanding of MRI image quality in the range [0, 1], offering clearer insights into the generated image quality.

Finally, we evaluated the quality of generated images for each distortion type using our proposed model to gain insights into improving generative models for higher-quality outputs. Table \ref{table-test-fake2} shows the results of this experiment, which exhibits varying levels of quality issues across different datasets with contrast being a common problem. Noise is predominant in generated samples from ADNI, while sampled images from BraTS and SynthRAD often suffer from blur and ringing effects, respectively. These findings suggest potential enhancements in generative models, such as adjusting loss functions to penalize specific artifacts.

\begin{table*}[!t]
\centering
\mysizeb
\caption{Comparison of quality assessment scores (mean$\pm$SD) using different methods across various synthetic datasets.}
\label{table-test-fake1}
\renewcommand{\arraystretch}{1.2}
\centering
\begin{tabular}{lcccccc}
\toprule
 & Proposed & 1 - PIQE & SSIM & BRISQUE & PSNR & FID \\
\bottomrule
ADNI & 0.871$\pm$0.017 & 0.392$\pm$0.028 & 0.670$\pm$0.014 & 43.364$\pm$1.556 & 17.346$\pm$0.505 & 196.010 \\
BraTS & 0.917$\pm$0.013 & 0.315$\pm$0.042 & 0.896$\pm$0.007 & 47.574$\pm$0.209 & 19.605$\pm$1.260 & 157.566 \\
SynthRAD & 0.891$\pm$0.019 & 0.466$\pm$0.055 & 0.727$\pm$0.022 & 46.719$\pm$1.081 & 17.736$\pm$0.860 & 191.584 \\
\toprule
\end{tabular}
\end{table*}

\begin{table*}[!t]
\centering
\mysizeb
\caption{Comparison of the quality scores (mean$\pm$SD) for different distortions estimated using the proposed method in the generated test samples.}
\label{table-test-fake2}
\renewcommand{\arraystretch}{1.2}
\centering
\begin{tabular}{lcccccc}
\toprule
 & Contrast & Bias & Ring & Ghost & Noise & Blur \\
\bottomrule
ADNI & 0.681$\pm$0.072 & 0.928$\pm$0.010 & 0.906$\pm$0.019 & 0.988$\pm$0.005 & 0.780$\pm$0.040 & 0.944$\pm$0.019 \\
BraTS & 0.845$\pm$0.011 & 0.969$\pm$0.003 & 0.957$\pm$0.011 & 0.976$\pm$0.004 & 0.916$\pm$0.053 & 0.837$\pm$0.019 \\
SynthRAD & 0.688$\pm$0.081 & 0.942$\pm$0.011 & 0.881$\pm$0.037 & 0.972$\pm$0.020 & 0.922$\pm$0.073 & 0.944$\pm$0.021 \\
\toprule
\end{tabular}
\end{table*}

\begin{figure*}[!t]
\centering
\begin{subfigure}[t]{0.99\textwidth}
\raisebox{-\height}{\includegraphics[width=0.99\textwidth, height=0.15\textheight]{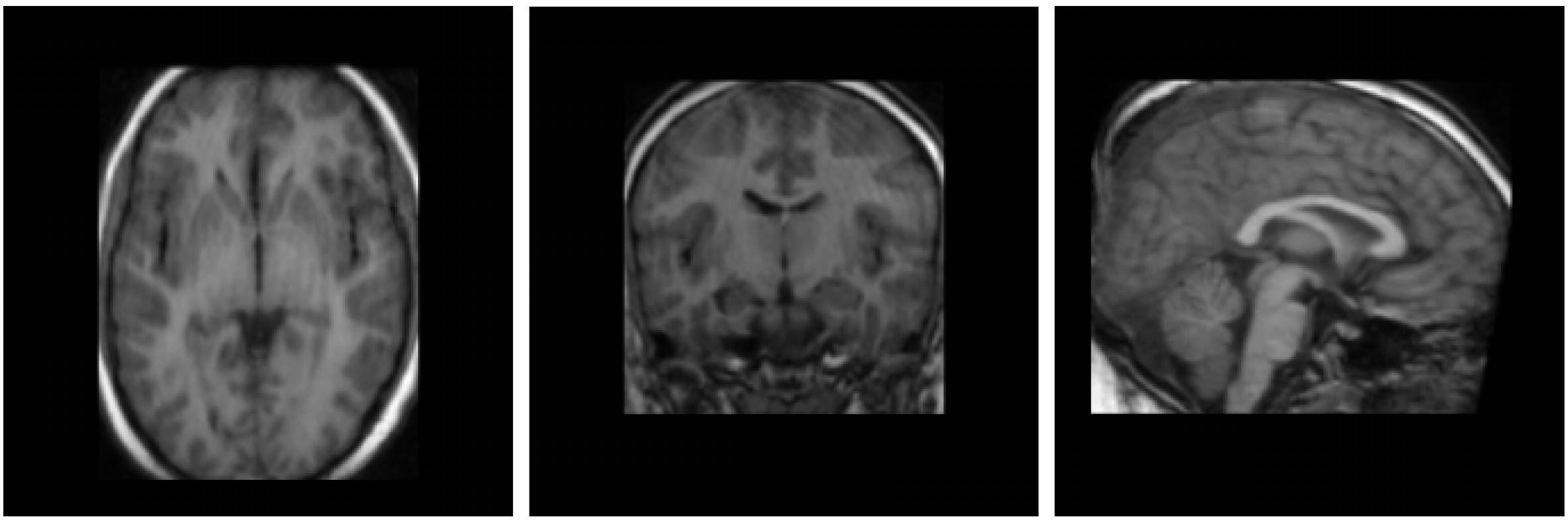}}
\end{subfigure}
\begin{subfigure}[t]{0.99\textwidth}
\raisebox{-\height}{\includegraphics[width=0.99\textwidth, height=0.15\textheight]{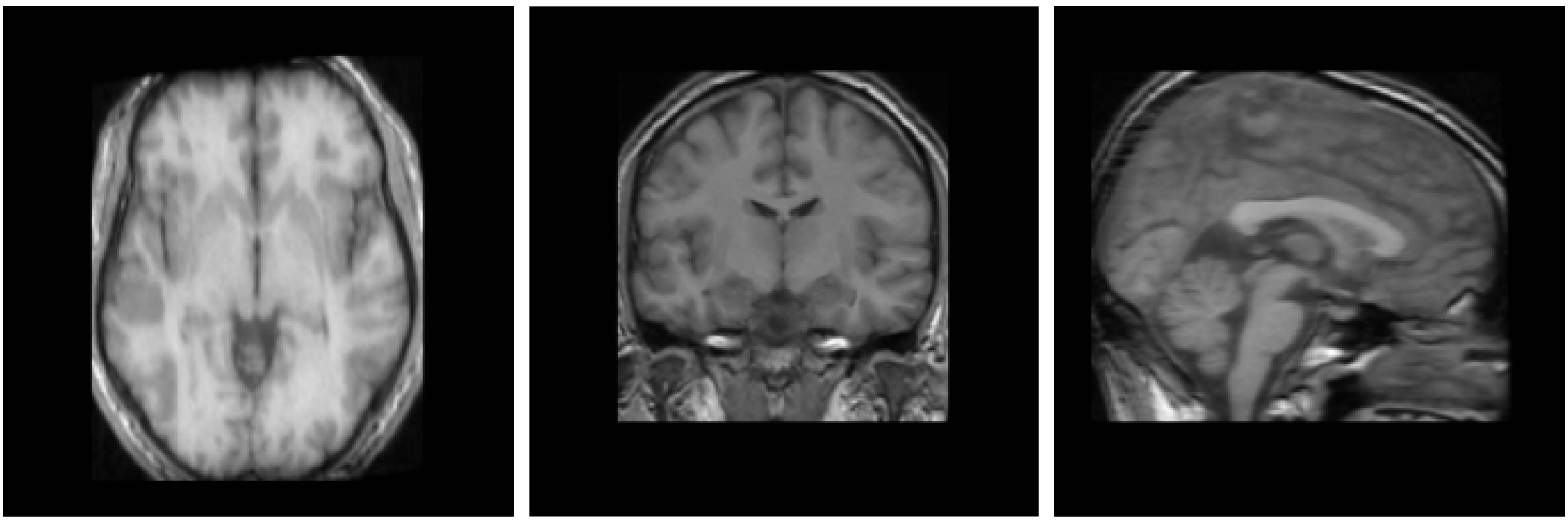}}
\end{subfigure}
\begin{subfigure}[t]{0.99\textwidth}
\raisebox{-\height}{\includegraphics[width=0.99\textwidth, height=0.15\textheight]{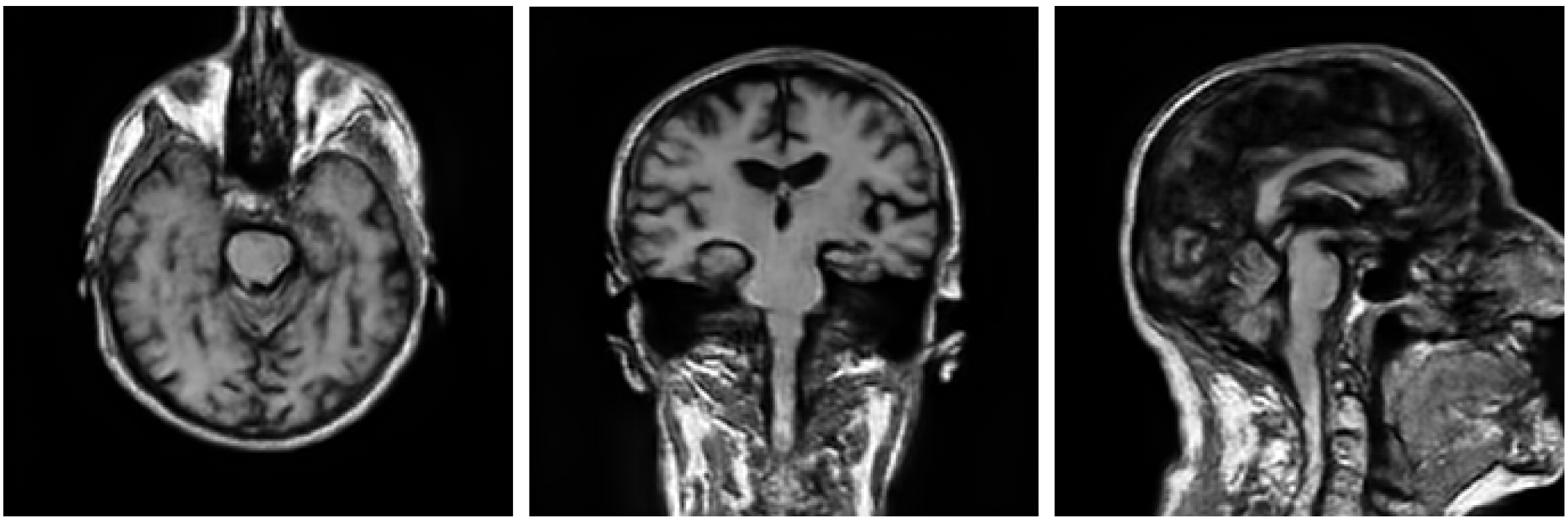}}
\end{subfigure}
\begin{subfigure}[t]{0.99\textwidth}
\raisebox{-\height}{\includegraphics[width=0.99\textwidth, height=0.15\textheight]{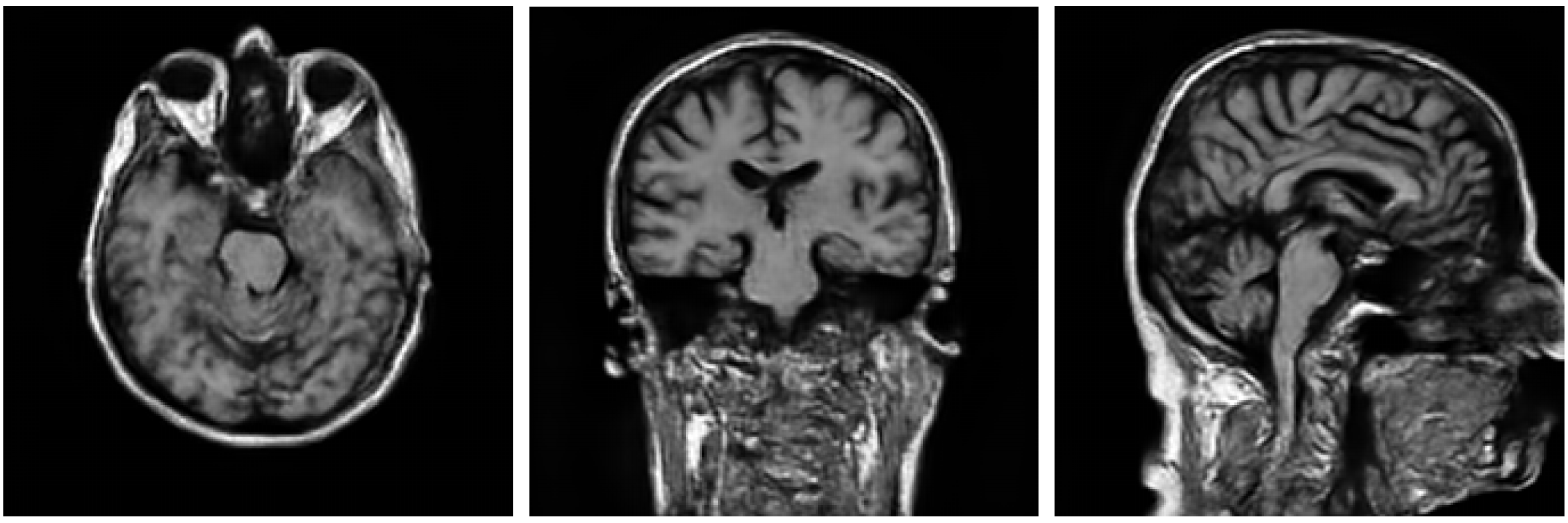}}
\end{subfigure}
\caption{Multiple views (axial-left, coronal-center, sagittal-right) of low-quality samples from real IBSR and synthetic ADNI datasets. Row 1: IBSR scan with ringing score 0.863. Row 2: IBSR scan with blur score 0.798. Row 3: ADNI scan with contrast score 0.580. Row 4: ADNI scan with noise score 0.687.}
\label{fig_artifact}
\end{figure*}

To evaluate image quality qualitatively, we examined samples from both real ISBR and generated ADNI datasets, noted for their lower quality scores in Tables \ref{table-test-real} and \ref{table-test-fake2}. Figure \ref{fig_artifact} illustrates these samples, demonstrating the proposed method's high accuracy in blind quality assessment.

\section{Conclusion}

In this study, we developed a 3D deep network for comprehensive assessment of MRI image quality, focusing on six common artifacts. Several datasets were utilized for training, validation, and testing of the quality of images against state-of-the-art metrics. Our approach provides a robust evaluation from multiple perspectives, with scores ranging from 0 to 1, using real datasets and synthetic MRIs generated with a trained diffusion model on high-quality 3D brain MRIs.

Comparisons with state-of-the-art metrics demonstrated that our method accurately identifies and quantifies distortions in MRI images while offering a multifaceted view of image quality. The proposed quality assessment provides intuitive insights across diverse datasets without requiring reference images. This can facilitate a better understanding of the quality of complex 3D medical images and contribute significantly to improving the quality of generated images during deep generative model training.

\section{Acknowledgments}

This project has received funding from Lundbeck Foundation with reference number R400-2022-617 and Pioneer Centre for AI, Danish National Research Foundation, grant number P1.

\section*{Disclosure of Interests}

The authors have no competing interests in the paper.

\bibliographystyle{splncs04} 
\bibliography{references}

\newpage
\begin{subappendices}

\section{Datasets} \label{appendix-data}

\subsection*{ADNI}

The primary dataset used in this study is derived from the Alzheimer’s Disease Neuroimaging Initiative (ADNI), specifically ADNI1: Screening 1.5T, comprising images from 503 subjects at baseline with an average voxel resolution of 1 $\times$ 1 $\times$ 1.2 mm$^3$ per voxel. These images were preprocessed using FreeSurfer's cross-sectional pipeline, which included motion and bias field correction, affine registration to the MNI305 space, resampling to an isotropic resolution of 1 mm, and intensity normalization to the range [0, 255]. These distortion-free standard images serve as references for training the quality network.

\subsection*{OASIS}

This dataset originates from the first phase of the Open Access Series of Imaging Studies (OASIS), specifically the 20Repeats collection. It comprises 40 MRI scans from 20 patients acquired using a 1.5T scanner, with images having an isotropic resolution of 1 mm per voxel.

\subsection*{Hammers}

This dataset comprises MRI scans from 30 healthy subjects, obtained using a 1.5T scanner placed at the epilepsy MRI unit. The scans have an average isotropic spacing of 0.94 mm per voxel.

\subsection*{IBSR}

The Internet Brain Segmentation Repository (IBSR) dataset consists of MRI scans from 18 healthy subjects obtained at the Center for Morphometric Analysis, Massachusetts General Hospital. The scans have an average resolution of 0.94 $\times$ 0.94 $\times$ 1.5 mm$^3$ per voxel.

\subsection*{IXI}

The Information eXtraction of Images (IXI) dataset comprises 581 MRI scans obtained from normal healthy subjects at the Centre for the Developing Brain, involving three hospitals in London. The scans were acquired using 1.5T and 3T scanners, with an average resolution of 0.94 $\times$ 0.94 $\times$ 1.2 mm$^3$ per voxel.

\subsection*{SynthRAD}

The challenge dataset for Synthesizing Computed Tomography for Radiotherapy (SynthRad2023) includes 180 MRI images acquired from radiation oncology departments across three Dutch university medical centers. These images were obtained using 1.5T and 3T scanners, featuring varying resolutions ranging from 0.22 mm to 1.12 mm per pixel.

\subsection*{BraTS}

The Brain Tumor Segmentation (BraTS) 2021 challenge dataset comprises 1,252 MRI scans of brain tumors collected from various institutions, scanners, and imaging protocols. The data is registered to the SRI24 anatomical template, resampled to a unit isotropic resolution, and has undergone skull-stripping preprocessing.

\end{subappendices}

\end{document}